\documentclass[prl,aps,superscriptaddress,groupedaddress,twocolumn]{revtex4}

\usepackage{xspace,amsmath,amsfonts,amsthm,amssymb,amsbsy}
\usepackage{graphicx}
\usepackage{amssymb,subeqnarray}
\usepackage{color}

\definecolor{pink}{rgb}{1,0.078,0.57}

\definecolor{green}{rgb}{0,0.7,0.9}

\newcommand{\ket}[1] {\left\vert #1 \right\rangle}

\newcommand{\dg}{^{\dagger}}

\newcommand{\vac}{\ket{v}}

\newcommand{\bP}{\mathbf{P}}

\newcommand{\bQ}{\mathbf{Q}}

\newcommand{\bR}{\mathbf{R}}

\newcommand{\fD}{\mathcal{D}}

\graphicspath{{../images/}}

\begin{document}

\title{Optical signature of quantum coherence in fully dark exciton condensates}

\author{Shiue-Yuan Shiau}
\affiliation{Physics Division, National Center for Theoretical Sciences, Hsinchu, 30013, Taiwan}
\author{Monique Combescot}
\affiliation{Sorbonne Universit\'e, CNRS, Institut des NanoSciences de Paris, 75005-Paris, France}
\date{\today}

\begin{abstract}
We predict that the collision of two fully dark exciton condensates  produces  bright interference fringes. So, quite surprisingly, the collision of coherent dark states  makes light. This remarkable effect, which is many-body in essence, comes from the composite boson nature of excitons, through the fermion exchanges they can have which transform dark states into bright states. The possibility of optically detecting quantum coherence in a regime where the system is hidden by its total darkness, was up to now considered as hopeless.
\end{abstract}

\maketitle

\begin{figure}[t]
\centering
\includegraphics[trim=0.7cm 1cm 0cm 1.3cm,clip,width=3in] {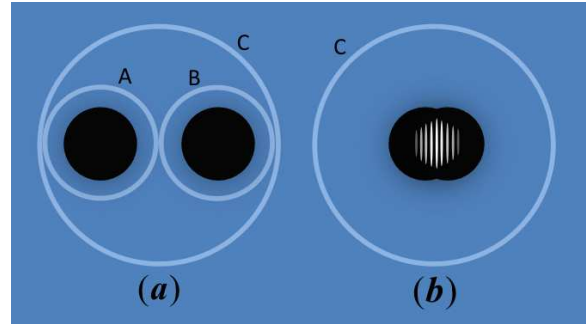}
\caption[]{\small (a) A dark condensate prepared in a large electrostatic C trap is split into two (dark circles) by turning on  the A and B traps. When the (A,B) traps are turned off, the two dark condensates move toward the C trap center and interfere. (b) Schematic view of the effect we predict: bright interference fringes appear in the middle of the dark region, as a striking signature of coherence in the colliding  fully dark condensates.}
\label{fig:1}
\end{figure}

This Letter aims to  write down the final act of the half-century long drama on exciton Bose-Einstein condensation (BEC), namely, how to evidence the condensate wave function coherence by an optical mean when the condensate is fully dark. The {\it tour de force} we propose relies on  the fact that carrier exchange between excitons couples dark to bright states. We use this many-body effect, exclusive to the composite boson nature of excitons, to here predict that the collision of two fully dark condensates must produce bright interference fringes that lead to a photoluminescence emission from an otherwise optically dark region (see Fig.~\ref{fig:1}). This effect constitutes the utmost evidence that the wave functions of the two colliding dark condensates are   quantum coherent. To better grasp the importance of this challenge, let us first recall the previous acts of the exciton condensation drama.

\textbf{\textit{The past}}---The quest for  BEC of semiconductor excitons---composite bosons (cobosons) made of one conduction electron and one valence hole---started in 1962 \cite{Blatt,Moskalenko,Keldysh1968}. The critical density and temperature for BEC being easy to  reach due to light carrier masses, excitons were for a long time thought to be the best candidate to experimentally produce this striking bosonic quantum effect. Yet, impressive progress in laser cooling has turned the tide and in 1995 the first ever BEC was realized in  $^{87}$Rb \cite{Anderson1995sci}, $^{23}$Na \cite{DavisPRL1995}, and $^7$Li \cite{BradleyPRL1995} atom gases.

As physicists were not understanding why exciton condensation eluded luminescence measurements, they  turned to polariton, which is a linear combination of one elementary boson, the photon, and one composite boson, the exciton. However, in  microcavities where experiments were performed \cite{RMPpolariton}, the photon component in the polaritons that condense is  large and thus quite different from a genuine  exciton. So, the problem of exciton BEC remained open. 

The difficulty with exciton condensation is the complexity of the exciton physics, many aspects of which have to be pieced together in order to possibly observe this quantum effect. 


First, to avoid  density collapse as well as condensate fragmentation into different momentum states \cite{Nozieres,monique2008prb}, a repulsive interaction must exist between excitons. Since two excitons can bind into a biexciton molecule when their excitonic dipoles point in  opposite directions, a way to avoid molecule formation is to use carriers located in spatially separate planes, as first proposed by Lozovik and Yudson \cite{Lozovik1975}. This bilayer geometry is particularly attractive because ``dipolar excitons" have a long lifetime that  makes possible the  study of cold exciton gases at  thermodynamical equilibrium.
Recent experiments \cite{PRL2011,APL2018,Nature2012} on exciton BEC have followed this idea \cite{Fnote1}.\

 Next, due to  sizable spin-orbit splitting  in the  degenerate GaAs upper valence band and to quantum well confinement \cite{Ivchenkobook}, the hole states with lowest energy  are characterized by  a quantum index commonly called  ``spin" $S_h=\pm3/2$, while the conduction  electrons  are simply characterized by their spin $S_e=\pm1/2$. Because electron-photon interaction conserves the  genuine spin $s=\pm1/2$,  electron-hole pairs coupled to photon, that are ``bright",  are such that  $S=S_e+S_h=\pm1$, while for ``dark" pairs not coupled to photon,  $S=\pm2$ \cite{MonicPRB2019}. Since Coulomb interaction also conserves the genuine  spin, bright excitons constructed on bright electron-hole pairs suffer interband Coulomb processes, while the  dark excitons  don't. This (repulsive) interband Coulomb interaction pushes the bright excitons up in energy; so, the lowest states which are the ones that condense, are dark \cite{Monique2007}. This allows associating  exciton condensation with a darkening of the luminescence when the temperature decreases \cite{alloing}, provided  the dark-bright energy splitting is small compared to the thermal energy, as for GaAs bilayer, whose splitting is estimated to be $\sim20\mu$eV, that is, ten times smaller than the thermal energy at 1 Kelvin.\

\begin{figure}[t]
\centering
\includegraphics[trim=4.2cm 8.7cm 4.8cm 5.5cm,clip,width=3.6in] {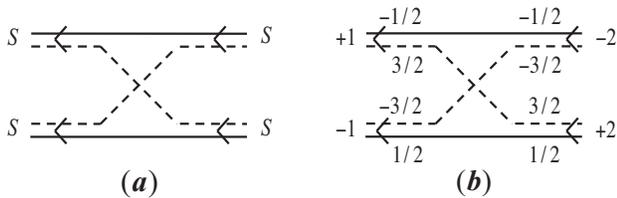}
\caption[]{\small  Hole exchanges between two
 dark $(\pm2)$ and/or bright $(\pm1)$ excitons, visualized through Shiva diagrams \cite{MoniqPhysreport,MoniqueSeanbook}. Electrons are represented by solid lines, holes by dashed lines, and excitons by electron-hole double lines. Because  exchanges conserve  the carrier spin, excitons resulting from exchange between same  $S=(\pm1,\pm2)$ excitons keep this $S$, as  in (a). By contrast, exchange between dark  excitons having opposite-$S$ leads to opposite-$S$ bright excitons, as  in (b), while exchange between a dark and a bright exciton would lead to the same dark and  bright excitons \cite{SM}.} 
\label{fig:2}
\end{figure}

 Yet, the indisputable signature for condensation  is the macroscopic coherence of the condensate wave function. It is clear that the dominant role played by the optically inactive states constitutes a severe constraint to experimental evidence, because the phase coherence of a fully dark condensate seemed optically unreachable. This impasse could only be unraveled through a deep understanding of the exciton composite nature and the interplay of their spin and orbital degrees of freedom. Excitons result from photon absorption; so, excitons are created in a  bright state by construction. Yet, being made of indistinguishable fermions, carrier exchanges  between excitons can transform two bright states into two dark states, and vice versa (see Fig.~\ref{fig:2}). As a result, (\textit{i})  although excitons are created in a bright state, dark excitons do exist in the system; (\textit{ii}) through exchange interaction, unimportant in the very dilute limit, the dark condensate acquires a coherent bright component above a density threshold and turns ``gray" \cite{Roland2012}. Up to now, optical access to  coherence has been possible in the gray regime:  it is through  the photoluminescence emitted by its bright component that spatial and temporal coherences of the exciton condensate have been measured \cite{dubin2017,Romain2018NJP}, and superfluidity observed from the formation of vertices \cite{Romain2018NJP,Dang2019}. \

Independently, Rapaport {\it et al} \cite{Rapaport}  reported on a fully dark condensation of dipolar excitons, at densities well above the limit for which a gray  condensate is energetically favorable, which is somewhat strange. It was argued  that dipolar repulsion between excitons stabilizes the fully dark condensate, and inhibits a coherent introduction of  bright excitons which  prevents the dark condensate from turning gray. As no evidence of quantum coherence has been shown in this fully dark system, this point seems moot. The optical effect we here predict will not only allow probing BEC at density too low for the condensate to be gray, but also provide a way to prove that dark BEC does occur in the experiments reported in \cite{Rapaport,Rapaport2,Beian2017epl}.

 \begin{figure*}[t]
\includegraphics[trim=0.8cm 5.5cm 0.7cm 4.5cm,clip,width=1.6\columnwidth]{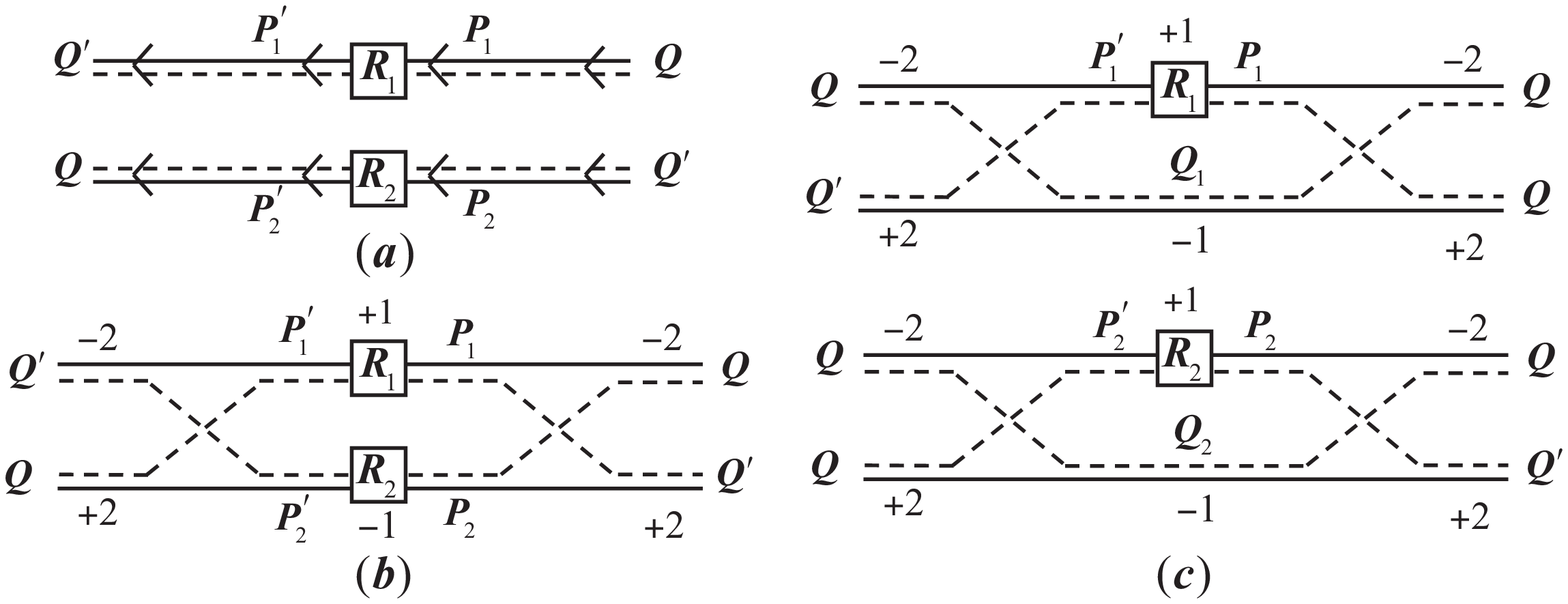}
\caption{\small (a) This diagram shows the process leading to Eq.~(\ref{1m}), obtained for elementary bosons, or for cobosons in the absence of fermion exchange. (b) When exchange is introduced, the excitons observed at $(\bR_1,\bR_2)$ positions are bright, but this process does not lead to interference.  (c) Disconnected diagram for hole  exchanges leading to the $m=1$ mode of Eq.~(\ref{Ameta'}). It involves three excitons from the $\bQ$ condensate and one exciton from the $\bQ'$ condensate. A similar exchange with two excitons $\bQ$ and two excitons $\bQ'$ leads to the $m=2$ mode.  The product of operators $B\dg_{\bR}B_{\bR}$ is visualized by a $\bR$ ``box''. These operators located at $\bR$ are linked to operators  for incoming and outgoing excitons having momenta $\bP$ and $\bP'$ through Eq.~(\ref{BR1_density}).    \label{fig:3} }
\end{figure*}

\textbf{\textit{Physics of the predicted effect}}---The interference pattern resulting from the collision of two condensates has been observed in the case of cold atoms \cite{Andrew1997,Vogels2003,Wang2005,Shin2005}. The  probability for detecting two elementary bosons located at $(\textbf{R}_1,\textbf{R}_2)$ is proportional to the two-boson spatial correlation function which has an oscillatory part in \cite{Javanainen1996a}
\begin{equation}  \label{1m}
nn'\cos \big((\textbf{Q}-\textbf{Q}')\cdot(\textbf{R}_1-\textbf{R}_2)\big)\,,
\end{equation}
 where $(n,\textbf{Q})$ and $(n',\textbf{Q}')$ are the densities and momenta of the two condensates. This effect comes from the process  of Fig.~\ref{fig:3}(a): a $\textbf{Q}$ boson and a $\textbf{Q}'$ boson are respectively observed at $\textbf{R}_1$ and $\textbf{R}_2$. Being indistinguishable, the bosons which leave $\textbf{R}_1$ and $\textbf{R}_2$ can as well be $\textbf{Q}'$ and $\textbf{Q}$ bosons. This  indistinguishability produces fringes with an oscillation characterized by the condensate momentum difference $(\textbf{Q}-\textbf{Q}')$. The $nn'$ factor comes from the number of ways to choose the $\bQ$ and $\bQ'$ bosons  among $N$ and $N'$.

 Due to their very small size, atoms  behave as elementary bosons, and  the only process that can happen is  the one of Fig.~\ref{fig:3}(a) \cite{Andrew1997}. Indeed, fermion exchanges between cobosons occur within their relative-motion volume $a^D$, while the coboson center of mass is delocalized over the  whole sample volume $L^D$, where $a$ is the coboson Bohr radius and $D$ the space dimension; so, these exchanges are controlled by the dimensionless many-body parameter 
 \begin{equation}
 \eta=N \Big( \frac {a}{L}\Big)^D=n\,a^D\,.
 \end{equation}
   For atoms, $\eta$ is close to zero while for excitons, whose size is much larger, $\eta$ can be  sizable; this explains why effects coming from carrier exchanges can be experimentally seen in excitonic systems.

 We can add exchange to the diagram of Fig.~\ref{fig:3}(a) by connecting its two parts; we get the diagram of Fig.~\ref{fig:3}(b).  If these exchanges occur between dark excitons having opposite spins, the excitons observed at $\textbf{R}_1$ and $\textbf{R}_2$  are bright. However, this ``connected" diagram \cite{disconnected} does not bring interference. The reason is that momentum conservation for the  incoming $(\textbf{P}_1,\textbf{P}_2)$ and outgoing $(\textbf{P}'_1,\textbf{P}'_2)$ bright excitons 
  \begin{equation}\label{mc1}
  \textbf{Q}+\textbf{Q}'=\textbf{P}'_1+\textbf{P}'_2=\textbf{P}_1+\textbf{P}_2
   \end{equation}
    leads to $\textbf{P}_1-\textbf{P}_1'=\textbf{P}_2'-\textbf{P}_2$, which can take any  value \cite{SM}. Interferences come from disconnected diagrams like the one of Fig.~\ref{fig:3}(c). Momentum conservation 
     \begin{subeqnarray}\label{mc2}
    \textbf{Q}+\textbf{Q}'=\textbf{P}'_1+\textbf{Q}_1=\textbf{P}_2+\textbf{Q}_2
     \\
      2\textbf{Q}=\textbf{P}_1+\textbf{Q}_1=\textbf{P}'_2+\textbf{Q}_2
       \end{subeqnarray}
    then gives $\textbf{Q}-\textbf{Q}'=\textbf{P}_1-\textbf{P}'_1=\textbf{P}'_2-\textbf{P}_2$. This process leads to bright excitons at $(\textbf{R}_1,\textbf{R}_2)$ due to the carrier exchanges it contains, and an oscillating correlation function in $\cos \big((\textbf{Q}-\textbf{Q}')\cdot(\textbf{R}_1-\textbf{R}_2)\big)$,  which manifests as bright fringes (Fig.~\ref{fig:1}(b)).

 \textbf{\textit{Experimental proposal}}---Indirect excitons formed in GaAs bilayer provide a suitable platform to show the effect we predict. Recent experiments \cite{Beian2017epl} have reached exciton density as large as $5\times 10^{10}$cm$^{-2}$  in a trap with diameter $10 \mu$m and  potential depth $5$meV, which corresponds to a many-body parameter $\eta \simeq 0.2$ for an interlayer separation that leads to excitons  having a Bohr radius $\simeq 20$nm. \

The observation of photoluminescent interference fringes from dipolar exciton condensates, however, faces the very low optical activity of these excitons. It is then necessary to repeat the experiment a few million times in order to obtain a sufficient signal-to-noise ratio. When the relative phase of the two fully dark condensates is random, the resulting interference patterns will have bright fringes  at different positions, and the averaging over these repeated experiments will blur the  fringe pattern into a bright spot. This phase randomness is avoided when the two colliding condensates come from the same source as proposed below, in analogy to the double-slit experiment.

What we suggest is somewhat similar to the procedure used  for cold atoms \cite{Andrew1997}: we first load the C trap (Fig.~\ref{fig:1}(a)) with excitons at a temperature above the one for BEC, in order to ensure that the trap contains  excitons. We then cool down the C trap until no light is emitted from it and we turn on the  A and B traps to split the dark condensate into two,  while keeping the excitons in the trap potential ground state. By suddenly removing the A and B traps, the C  trap that encompasses the two condensates, exerts a force which  pushes them toward its center with momenta $\textbf{Q}$ and $\textbf{Q}'=-\textbf{Q}$, where they interfere. Interference fringes are formed in the central region of the C trap: dark fringes, similar to the ones for atoms, exist but they cannot be seen.  Bright fringes are also formed that  can be optically detected. To produce bright fringes of  width $\sim\mu$m, the C trap must be rather shallow compared to the depths of the A and B traps:  indeed, a fringe width of $1\mu$m corresponds to an exciton momentum $h/\mu$m, that is, a kinetic energy $\sim3\mu$eV, which is provided by the C trap potential.

\textbf{\textit{Mathematical support}}--
The interference pattern of two colliding condensates made of excitons with momenta $(\bQ,\bQ')$ and densities  $(n,n')$, is obtained from the spatial correlation of two excitons located at $(\bR_1,\bR_2)$, these excitons having to be bright in order to be optically detected. The spatial correlation function for two bright excitons having circular polarizations $(\sigma_1,\sigma_2)$ with $\sigma_{i}=\pm1$, is given by
\begin{equation}
\langle B\dg_{\bR_1;\sigma_1}  B\dg_{\bR_2;\sigma_2} B_{\bR_2;\sigma_2}  B_{\bR _1;\sigma_1} \rangle \,,\label{SNN'2}
\end{equation}
 the expectation value being taken in the two-colliding-condensate state $(N,\textbf{Q};N',\textbf{Q}')$. 
 

\noindent${\it Case~1}$: If the two condensates were in the same bright state $\sigma$, Eq.~(\ref{SNN'2}) would give a spatial correlation  that oscillates just as for elementary bosons \cite{Javanainen1996a} in Eq.~(\ref{1m}), provided  $\sigma_1=\sigma_2=\sigma$. Indeed, when exchanges occur between same $S$ excitons, this index  does not change (see Fig.~\ref{fig:2}(a)). \
 


 
\noindent${\it Case~2}$: Likewise, carrier exchanges between same-$S$ dark excitons do not produce bright excitons; so, the bright exciton destruction operator $B_{\bR;\sigma}$ acting on same-$S$ dark condensates readily gives  zero whatever $\sigma$: correlations do exist, but they cannot be optically detected.
 
\noindent${\it Case~3}$: If one condensate contains $(+2)$ excitons only and the other  $(-2)$ excitons only, carrier exchange between these opposite-$S$ dark excitons produces bright excitons (Fig.~\ref{fig:2}(b)). The two-bright-exciton correlation function differs from zero but no  fringes are produced. Indeed, for bright fringes to appear, it is  necessary  to have a macroscopic amount of $(+2)$ and $(-2)$ excitons in one condensate at least, this condensate being either unpolarized $(B\dg_{2})^{N_+}(B\dg_{-2})^{N_-}\vac$ or polarized $(g_2B\dg_{2}+g_{-2}B\dg_{-2})^N\vac$: the appearance of bright fringes indeed  is a quite subtle many-body effect! Actually, through mean-field theory, it has been shown \cite{Monique2007} that the dark exciton condensate is polarized with $|g_2|=|g_{-2}|$, thus making the formation of bright fringes {\it a priori} possible.\

  For a state made of two polarized dark condensates  
\begin{equation}
\ket{\psi_{{N},{N'}}} = (\fD\dg_\bQ )^{{N}}({\fD}\dg_{\bQ'} )^{{N'}} \vac \label{2darkconde}
\end{equation}
with $\fD\dg_\bQ=g_2B\dg_{\bQ;2}+g_{-2}B\dg_{\bQ;-2}$, we find that to the lowest order in density, the  correlation function (\ref{SNN'2}) oscillates with two modes, 
\begin{equation}
|g_2g_{-2}|^4\Lambda^4\sum_{m=(1,2)}  A^{(m)}_{\eta,\eta'}\cos \Big(m(\bQ-\bQ')\cdot(\bR_1-\bR_2)\Big)\label{Ameta'}
\end{equation}
whatever $(\sigma_1,\sigma_2)$. The  dimensionless coefficient $\Lambda^2$ contains two fermion-exchange scatterings, $A^{(m)}_{\eta,\eta'}$  being equal to $8\eta\eta'(\eta+\eta')^2$ for $m=1$, and to $2(\eta\eta')^2$ for  $m=2$. Higher $m$ modes do exist but they are of higher order in density.

 To obtain this result, we used the composite boson  many-body formalism \cite{MoniqPhysreport,MoniqueSeanbook}, which allows handling fermion exchanges between cobosons in an exact way; its detailed derivation, including the calculation of $\Lambda^2$ in the case of GaAs bilayer, is given in the Supplements \cite{SM}.  The salient points about Eq.~(\ref{Ameta'}) are:

 \noindent \textbf{\textit{(i)}} the $g_2g_{-2}$ factor: it proves that  the two types of dark excitons are necessary  to produce a non-zero  bright exciton correlation. This correlation is entirely due to fermion exchanges between $(+2,-2)$ excitons, in the absence of Coulomb process; so, the associated exchange scattering that enters $\Lambda^2$ is  dimensionless, unlike usual energy-like interaction scatterings.  
 
\noindent \textbf{\textit{(ii)}} the $(\eta\eta')^m$ density dependence in  the $m$ mode and the $(\bQ-\bQ')$ difference: they show that the two condensates join together to produce  oscillations, as for elementary bosons. Moreover, the existence of bright fringes supports a polarized  BEC because they require two types of dark excitons with same momentum.

\noindent \textbf{\textit{(iii)}} the higher oscillatory modes  $m=(2,3,...)$, while  elementary bosons only have the $m=1$ mode (Eq.~(\ref{1m})).


  Let us now go somewhat  deeper into the calculation. The creation operator of a bright exciton $\sigma$ located at $\bR$ is related to creation operators for excitons  having a center-of-mass momentum $\bP$ through $B\dg_{\bR;\sigma}=\sum_{\bP}B\dg_{\bP;\sigma}\langle \bP|\bR\rangle$
with  $\langle \bP|\bR\rangle=e^{-i \bP\cdot \bR}/L^{D/2}$; so, 
\begin{equation}\label{BR1_density}
 B\dg_{\bR_1;\sigma}B_{\bR_1;\sigma}=\frac{1}{L^D}\sum_{\bP'_1\bP_1}e^{i\bR_1\cdot(\bP_1-\bP'_1)}B\dg_{\bP'_1;\sigma}B_{\bP_1;\sigma}\,.
\end{equation}
We visualize this product of  operators  by  a $\bR_1$ box with an incoming bright exciton $\bP_1$ and an outgoing bright exciton $\bP'_1$ (see Fig.~\ref{fig:3}). When  the excitons making the condensates are all dark as  in Eq.~(\ref{2darkconde}), the exciton composite nature must enter into play through fermion exchanges as in Fig.~\ref{fig:2}(b), in order for the $(\bP',\bP)$ excitons to be bright. Starting from the diagram of Fig.~\ref{fig:3}(a), the simplest way is to have the observed $(\bR_1,\bR_2)$ bright excitons resulting from a  hole  exchange between a dark exciton $(+2)$ from the $\bQ'$ condensate and a dark exciton $(-2)$ from the $\bQ$ condensate. This process, shown in Fig.~\ref{fig:3}(b), could be optically detected for $\sigma_1{=}-\sigma_2$, but not for $\sigma_1=\sigma_2$.  Momentum conservation given in Eq.~(\ref{mc1}) produces terms in  $e^{i(\bR_1-\bR_2)\cdot(\bP'_1-\bP_1)}$ through Eq.~(\ref{BR1_density}), but does not enforce $(\bP'_1-\bP_1)$ to be constant. So, it does not lead to fringes.


 The more complex process of Fig.~\ref{fig:3}(c)  also contains carrier exchanges between $(+2,-2)$ excitons, as necessary to have bright excitons  at $(\bR_1,\bR_2)$, but momentum conservation given in Eq.~(\ref{mc2}) now imposes  $\bP_1-\bP'_1=\bQ-\bQ'$, which leads to a constant phase in Eq.~(\ref{BR1_density}); and similarly $\bP_2-\bP'_2=\bQ'-\bQ$. So, we  end with a $\cos \big((\bQ-\bQ')\cdot(\bR_1-\bR_2)\big)$ term  by interchanging $(\bR_1,\bR_2)$, that is, a $m=1$ mode. Since this process involves three $\bQ$ excitons and one $\bQ'$ exciton, it  must appear with a $\eta^3\eta'$ density dependence. Moreover, since it involves two pairs of excitons $(+2,-2)$ on both sides, taken from the polarized  $\ket{\psi_{{N},{N'}}}$ condensate, it has to contain a $|g_2 g_{-2}|^4$ factor.  This, and the other two processes involving one $\bQ$ exciton and three $\bQ'$ excitons, and two pairs of $(\bQ,\bQ')$ excitons, yield the $m=1$ term in Eq.~(\ref{Ameta'}). Note that for the process of Fig.~\ref{fig:3}(c) to  produce an oscillation, the $\bQ$ condensate must contain $(+2)$ and $(-2)$ excitons, while the $\bQ'$  condensate can contain $(-2)$ excitons only. 


Another process, similar to the one of Fig.~\ref{fig:3}(c), but with two $\bQ$ excitons and two $\bQ'$ excitons on the same side, instead of $(\bQ,\bQ)$ and $(\bQ,\bQ')$, also has a $|g_2 g_{-2}|^4$ factor, but a density dependence in $(\eta\eta')^2$. Momentum conservation now imposes $2\bQ'=\bP'_1+\bQ_1$ and $\bP_1+\bQ_1=2\bQ$, which give $2(\bQ-\bQ')=\bP_1-\bP'_1=\bP'_2-\bP_2$; so, this process brings the $m=2$ term of Eq.~(\ref{Ameta'}). For the $m=2$ mode to appear, the two dark condensates have to both contain  $(+2)$ and $(-2)$ excitons.

\textbf{\textit{Discussion}}---To keep the relative phase of the two condensates fixed for each repeated experiment, we must allow the particle numbers of the two condensates to vary, since the system phase and the particle number are conjugate variables. Let us for simplicity  consider  that excitons are  elementary bosons interacting through an effective two-body potential $\xi^{eff}$. Because the condensate energy depends on  the number of particles it contains, the time evolution operator $e^{-iHt}$ is going to  produce different phases to the two evolving condensates having different particle numbers. Mean-field calculation \cite{Jav1997PRL} actually  shows that the diffusion (root-mean-square deviation) of the relative phase  increases with time $t$ as $\sqrt{N}\xi^{eff} t$,  the $\sqrt{N}$ dependence coming from the fluctuation in the $N$-particle binomial distribution of the two traps. So, if the time lapse between the condensate splitting and the detection is short, the interference pattern will stay the same in each experiment.

{\bf To conclude}, we propose an optical way to probe quantum coherence in an excitonic system hidden by its darkness. Such a signature seemed at first hopeless. The effect  we here propose is based on the fact that two opposite-spin dark excitons transform into bright excitons through carrier exchange scatterings, that are dimensionless. The bright fringes we predict from the collision of two fully dark  exciton condensates constitute the utmost evidence of  coherence in these hidden states.


We thank Fran\c cois Dubin and Yia-Chung Chang for fruitful discussions and constructive suggestions during the  revision of our manuscript. We also thank You-Lin Chuang for suggestions on the experimental realization of the effect we predict.


\begin{thebibliography}{99}
\bibitem{disconnected} By disconnected diagrams, we mean that the $\bR_1$ and $\bR_2$ boxes belong to two separate Shiva diagrams, while for  connected diagrams, these two boxes are in a Shiva diagram.
\end{thebibliography}
\end{document}